\def\nii{[N~{\sc ii}]}   
\def\sii{[S~{\sc ii}]}   
\def\oiii{[O~{\sc iii}]}       
\def\ha{H$\alpha$}     
\def\hb{H$\beta$}       
\def\ii{{$i^\prime$}}      
\def\rr{{$r^\prime$}}      
\def\gg{{$g^\prime$}}      
\def\heii{He~{\sc ii}}       
\def\arcsec{\hbox{$^{\prime\prime}$}}        

\documentclass{iaus} 
\usepackage{graphicx} 
 
\title[LG surveys for PNe] %% give here short title %% 
{Local Group surveys for Planetary Nebulae} 
 
\author[L. Magrini]   %% give here short author list %% 
{Laura Magrini$^1$ 
} 
 
\affiliation{$^1$INAF--Osservatorio Astrofisico di Arcetri,  
Firenze, Largo E. Fermi, 5, 50125 
Italy \break email: laura@arcetri.astro.it\\[\affilskip]} 
 
\pubyear{2006} 
\volume{234}  %% insert here IAU Symposium No. 
\pagerange{} 
\date{} 
\jname{Planetary Nebulae in Our Galaxy and Beyond} 
\editors{M. J. Barlow \& R. H. M\'endez, eds.}
\begin{document} 
 
\maketitle 
 
\begin{abstract} 
The Local Group (LG) represents the best environment to study in 
detail the PN population in a large number of morphological types of 
galaxies. The closeness of the LG galaxies allows to investigate the 
faintest side of the PN luminosity function and to detect PNe also in 
the less luminous galaxies, the dwarf galaxies, where a small number 
of them is expected. 
 
A review of the results of the most recent imaging surveys in the LG 
is presented. 
%The relationships between the number of PNe and several 
%properties of the host galaxies, as for example the total luminosity, 
%the star formation history, and the metallicity, are analyzed. 
Some applications of the surveys for PNe to the study of the 
star formation history of the host galaxies are analyzed.    
In addition, these new observational data are an invaluable resource 
for follow-up spectroscopy to derive the chemical properties of not 
only PNe, but also other important emission-line sources like HII 
regions.  These are fundamental tools for the discussion of the 
chemical evolution of the host galaxies, mapping the history of their 
chemical enrichment at different epochs. The latest results on this 
subject are presented. 
\end{abstract} 
 
\firstsection % if your document starts with a section, 
\section{The Local Group of galaxies}

The Local Group (LG) is composed of about 40 members. 
Their belonging  to the LG is established using the  
following criteria (\cite{vdb00}): 
i) the distance to the galaxy should be $\leq$1.5 Mpc;  
ii) the galaxy should have a radial velocity which 
indicates that it might be be dynamically bound to the LG;  
iii) it should not appear to be associated with a group of 
galaxies that  is know  to be located well beyond the limits of the LG.  
 
The two more massive galaxies, the spiral galaxies M31 and Milky Way,
constitute the barycentres of two main subgroups.  The Andromeda
subgroup contains M31, its dwarf elliptical and spheroidal companions, 
among them the brightest ones are M32, NGC205, NGC185, NGC147, the
late-type spiral galaxy M33, and many other dwarf galaxies, as IC10,
DDO~210, SagDIG, Cas~dSph.  The Galaxy subgroup is composed by the
Milky Way, the two irregular galaxies LMC and SMC, and a large number
of dwarf spheroidal galaxies, like Draco, Ursa Minor, Fornax, Carina,
LeoI, LeoII, Sculptor, Sagittarius.  In addition, a number of isolated
galaxies are present, like the dwarf irregular galaxies NGC~6822,
IC~1613, WLM and Leo~A.  In the outskirts of the LG, a small
association of dwarf galaxies (including NGC~3109, Antlia, Sextans A
and B), represents the closest external group of galaxies. This group
is not dominated by giant galaxies.  Due to their proximity, the LG
galaxies have been  subject of deep and detailed studies (see the reviews by
\cite{vdb00}, \cite{mateo98}).  The structure of the LG, a small
cluster containing essentially all morphological types of galaxies, except giant
elliptical, is typical in the Local Universe. Consequently the results obtained
from these galaxies might be extrapolated and generalized for farther
galaxies.

\section{The search for PNe in the LG}\label{sec:surveys} 
 
\subsection{A brief history of the search for PNe in the LG} 
 
The first extragalactic (and LG) PN was discovered by Miss Annie
Cannon.  In her catalogue of peculiar spectra in the
Magellanic Clouds (\cite{cannon1933}), she also classified
emission-line stars with \oiii\ 5007, 4959 and \hb.  At first, these objects
were not identified as PNe, while later some of them were
recognized to be PNe.
  
The first efforts addressed  to the detection of PNe in the LG
were realized by \cite{b55} with the discovery of five PNe in M31 and
contemporaneously by \cite{l55} with 17 PNe in the SMC.
Afterward, during the 60s and the early 70s the search was limited to
the Magellanic Clouds, since the larger distances of the other galaxies
limited the investigations.  
A review by Westerlund (\cite{w68}) described the status at that time with 
45 PNe in LMC and 30 in SMC discovered by several investigators.  
Subsequently, in the 70s and 80s a series of papers, primarily by Ford,
Jacoby and collaborators, reached  other LG galaxies: NGC185,
NGC205, M32 (\cite{f73}), NGC147 (\cite{f77}), Fornax (\cite{d78}),
IC10, LeoA, Sextans A, Pegasus, and WLM (\cite{jl81}) and discovered new
PNe in M31 (\cite{fj78}), LMC and SMC (\cite{j80}).  Only
with the advent of CCD photometry, a large number of PNe was
discovered in M31 and in its companions (\cite{c89}). These authors 
used for the first time the  PN
luminosity function as a standard candle
of extragalactic distances.  
This issue represented the main topic of the study of extragalactic
PNe during the following decade, moving the attention from the LG to
the giant galaxies beyond the LG where a larger PN population could be
found.
During the first surveys for PNe, M33 was partially neglected.  
An objective-prism survey for emission-line objects was attempted by Lequeux et al.
(\cite{l87}), but the first deep survey for PNe was obtained only in 2000  
by Magrini et al. (\cite{m00}).  
At the present time,  M31 is the LG galaxy with the largest number of candidate PNe.
In fact a new instrument, dedicated to the study of extragalactic PNe,
the Planetary Nebula Spectrograph (\cite{douglas02}), allowed to
discover there 2615 PNe and at the same time to measure their radial
velocities (\cite{merrett06}).

\subsection{The surveys of LG galaxies $2001-2006$} 
 
Currently, a new attention for the PN population in the LG has revived. 
During  the last five years, two main surveys of the LG galaxies 
have been carried out: the {\bf Local Group Survey} (P.I. P. Massey) and the  
{\bf Local Group Census} (P.I. N. Walton).  
 
The main aim of the 3-year survey Local Group Survey ({\bf LGS}) had been to
study the galaxies currently active in star-formation in the LG
(\cite{massey02}).  The galaxies observed by the LGS
are (in brackets the number of fields when greater than one): M31(10),
M33(3), NGC 6822, IC 1613, IC 10, WLM, Pegasus, and Phoenix, and also
Sextans A and Sextans B, located in the outskirts of the LG.  These
galaxies have been observed with he CTIO and KPNO 4-m telescopes
equipped with the wide-field mosaic cameras (a mosaic of 8k$\times$8k CCDs) using
broad and narrow-band filters: UBVRI, \ha, \sii\,671.7,673.1~nm, and
\oiii\ 500.7~nm.  More information about this survey, the
publications, the data release, can be found at:
 
 \centerline{\tt http://www.lowell.edu/users/massey/lgsurvey.html}

The Local Group Census ({\bf LGC}) (2001-2003, 45 nights) has provided deep
broad- and narrow-band images of LG galaxies with Dec$\ge-30^\circ$
(\cite{cm05}), with the important exception of M~31, for which
observations aimed at detecting PNe had already been obtained by
other ING programs.  
Observations were carried out using the Wide
Field Camera (WFC) of the 2.5m~INT telescope of the Isaac Newton Group
of Telescopes  at La Palma. The WFC is equipped with a mosaic of
four 2k$\times$4k CCDs covering a field of view of 34$'$$\times$34$'$.
The pixel scale is 0.$\arcsec$33.  The galaxies observed by the LGC
are (in brackets the number of WFC fields): Draco (4), DD210,
EGB~0427+63, GR~8, IC~10, IC~1613, Leo A, Leo~II, Leo~I, M~33 (7),
NGC~147, NGC~185, NGC~205 (2), NGC~6822 (4), Pegasus, Sextans A,
Sextans B, Ursa Minor (4), and WLM.
 The majority of the  galaxies were observed in four narrow-band 
filters \oiii\ 500.7~nm, \ha+\nii, \sii\,671.7,673.1~nm, and 
\heii\ 468.6~nm.  
In addition, broad-band images were obtained using  the  Sloan 
\gg, \rr, \ii\  and the {\it Str\"omgren Y} filters. 
These filters have been used as a continuum for the narrow-band images  
to detect emission-line objects.  
In particular  the {\it Str\"omgren Y} and \gg\ filters were used  
for the  narrow-band filter \oiii\ 500.7~nm and \rr\ for  \ha\ + \nii.  
Typical exposure times were 60~min for the narrowband filters, 30~min for {\it 
Str\"omgren Y}, and 20~min for the Sloan filters.  
The completeness reached in the search  for unresolved emission--line 
objects in relatively crowded regions of galaxies has been  
approximately m$_{[OIII]} \sim$ 24.5~mag, where   
m$_{[OIII]}$ is defined by  Jacoby (\cite{j89}). 
More details about the survey and a complete list of the data obtained 
can be found at:  
 
\centerline{\tt http://www.ing.iac.es/$\sim$rcorradi/LGC/}

Both surveys leaded up to a significant increase of the number of  
candidate PNe in LG galaxies.  
From 2000 to the present year the number of candidate PNe identified in the LG 
has increased of 100\% from $\sim$1500 to $\sim$3000 (excluding the Magellanic 
Clouds and the Galactic PNe, see \cite{p06}).  
The data from the {\bf LGS} observations were used by Ciardullo et al. (\cite{ci04}) to identify  
152 PNe in M~33. 
The results obtained until now by the {\bf LGC} team in the search for PNe
are: Sextans B (5 PNe) (\cite{m02}), Sextans A (1), IC10 (16), LeoA (1)
(\cite{m03b}), NGC205 (35), NGC185 (5), NGC147 (9) (\cite{c05}), NGC6822
(17) (\cite{l05}), IC1613 (2), WLM (1), GR8 (0) (\cite{m05b}).  
In particular, the PNe discovered by the LGC represent the first step
in the study of the PN population in low-metallicity, dwarf irregular
galaxies, since so far the PN population in these galaxies
has been neglected.

\subsection{The identification techniques and criteria} 
 
The most common  techniques used to search for PNe in external galaxies are  
the on-band off-band technique and the colour-colour diagrams.
In addition, in wide intracluster regions automated detection algorithms 
have been directly applied to the continuum-subtracted frames (\cite{f03}). 

The {\bf techniques} used by the LGC are described in the following.  
In order to search for emission-line objects in the LG galaxies,   
we subtracted from the \oiii\ frames the properly scaled {\it Str\"omgren Y} or \gg\ frames,  
choosing for each galaxy the best quality continuum. 
Similarly, the \rr\ frames were subtracted from the \ha\ images.   
\ha$-$\rr\ {\it vs.} \oiii$-$\gg\  diagrams were also obtained.   
The colour-colour diagrams are very powerful  to detect unresolved 
sources in uncrowded sky regions, like the external 
regions of galaxies and the intracluster areas (\cite{arna2003}), 
while the continuum-subtracted images are proved  
to be more effective in crowded zones like the inner regions of galaxies 
(\cite{c05}). 
 
The  {\bf criteria} used in the imaging searches 
to identify a ``bona fide'' candidate PNe are the following:  
{\it i)} they must appear in both the \oiii\ and \ha+\nii\ images, but not in the 
continuum frames; 
{\it ii)} they must be unresolved at the distance  
of LG galaxies, taking into account the typical physical size of PNe (0.1-1~pc);
{\it iii)} they should show the excitation parameter  R=\oiii/\ha+\nii $>$ 
1 or somewhat larger (Magrini et al. \cite{m00}, \cite{c02})  to distinghuish between 
compact HII regions and PNe in a statistical sense (but see, for example, the Type 
I PN in Sextans A, \cite{m05a});
{\it iv)} the candidate PNe in the top 1 mag of the PNLF should have generally 
R$\geq$2 (see Fig. 2 of \cite{c02}).

Table~\ref{tab-pn} shows the galaxies with detected candidate PNe belonging to the LG and to its outer fringes.  
The morphological types, the
luminosities, the distances are from van den Bergh (2000).  
The references to the latest identification of PNe (N. PNe) and to their spectroscopic
confirmation (N. S.) are in the adjacent columns.
The search for PNe is almost complete within the
first 2-3 magnitude of their luminosity function for all the galaxies
where a PNe population is expected ($\log$L$_v$ $\geq$ 6.7, \cite{m03b}).
On the other hand, the spectroscopic confirmation has been obtained
for only 10\% of them (see Table\ref{tab-pn}).

\begin{table} 

\begin{center} 
\setlength\tabcolsep{5pt} 
\begin{tabular}{llrrrlrl} 
\hline\noalign{\smallskip} 
Name & T &  logL$_V$ & D  & \multicolumn{1}{c}{N. } & Reference& \multicolumn{1}{c}{N.  } & Reference \\ 
     &      &        & [kpc] & \multicolumn{1}{c}{PNe} &  & \multicolumn{1}{c}{S.} &                   \\ 
\noalign{\smallskip} 
\hline 
\noalign{\smallskip} 
%\multicolumn{2}{c}{\it Local Group} &  & & & \\ 
M31        & Sb      & 10.43 &   760 &   2615     & \cite{merrett06}      & 30  & \cite{jc99}   \\ 
M33        & Sc      &  9.51 &   795 &    152     & Ciardullo et al. \cite{ci04}& 26  & \cite{m03a} \\ 
LMC        & Ir      &  9.35 &    50 &   1000  & \cite{rp05}		  & 141 & \cite{l06a} \\ 
SMC        & Ir      &  8.79 &    59 &    132     & \cite{j05}  	  & 42  & \cite{l06a} \\ 
M32        & E2      &  8.55 &   760 &     46     & \cite{merrett06}      & 14  & \cite{r02} \\ 
NGC205     & Sph     &  8.51 &   760 &     35  	  & \cite{c05}  	  & 13  & \cite{r02}  \\ 
IC10       & Ir      &  8.47 &   660 &     16     & \cite{m03b}  	  & -   & \cite{m06}\\ 
NGC6822    & dIr     &  8.35 &   500 &     17     & \cite{l05}  	  & 17  & \cite{l06b},  \\
	   &         &	     &       &            &                       &     & \cite{her06}\\
NGC185     & Sph     &  8.19 &   660 &      5     & \cite{c05}  	  & 5   & \cite{r02} \\ 
IC1613     & dIr     &  8.07 &   725 &      2     & \cite{m05b}  	  & -   & \cite{c06}\\ 
NGC147     & Sph     &  7.99 &   660 &      9     & \cite{c05}            & 8   & \cite{g06} \\ 
WLM        & dIr     &  7.61 &   925 &      1     & \cite{m05b}           & -   & \\ 
Sagitt.    & dSp     &  7.47 &    24 &      4     & \cite{z06}            & 4   & \cite{z06} \\ 
Fornax     & dSp     &  7.19 &   138 &      1     & \cite{d78}            & 1   & \cite{d78} \\ 
Pegasus    & dIr     &  6.87 &   760 &      1     & \cite{jl81}           & -   & \\ 
LeoA       & dIr     &  6.55 &   690 &      1     & \cite{m03b}           & 1   & \cite{vz06} \\ 
%\hline 
\noalign{\smallskip} 

NGC3109    & dIr     &  8.27 &  1330 &     13  & \cite{l06b},             &12	 &\cite{l06b},  \\
	   &         &       &       &         & \cite{pena06}            &      &\cite{pena06}\\
SextansB   & dIr     &  7.63 &  1600 &      5  & \cite{m02}  	          &5     &\cite{m05a} \\ 
SextansA   & dIr     &  7.67 &  1320 &      1  & \cite{m03b}  	          &1     &\cite{m05a} \\ 
%GR 8       & dSph    &  6.59 &  2200 &      0  & \cite{m05b}              & -    &\\ 
\hline\\ 
\end{tabular} 
\end{center} 
\caption{The Local Group and outskirts galaxies with PNe detection.
The morphological types (T), luminosities ($\log L_{V}$ and distances (D)  are from van den Bergh (2000). The references to the number of candidate PNe (N.PNe) and to their spectroscopic 
confirmation  (N. S.) are in the adjacent columns. } 
\label{tab-pn} 
\end{table}

\section{What can we learn from Local Group PNe (and HII regions)?}\label{sec:results} 

Why LG PNe are so special and why do we want to study them in detail?
They have the advantage to be located outside the Galaxy, and consequently to
have well known distances. On the other hand they are relatively close.
Thus, among the extragalactic PNe, physical and chemical properties
can be derived measuring directly electron temperature and densities
only for them.
Furthermore, they belong to a variegated sample of galaxies with different 
morphology, metallicity, chemical evolution, and star formation history.
In the following sections, some examples of their application to 
the study of LG galaxies are described.

\subsection{Photometry}

\begin{figure}
\label{fig-sfr}
\centering
 \includegraphics[height=8cm,width=8cm,angle=0]{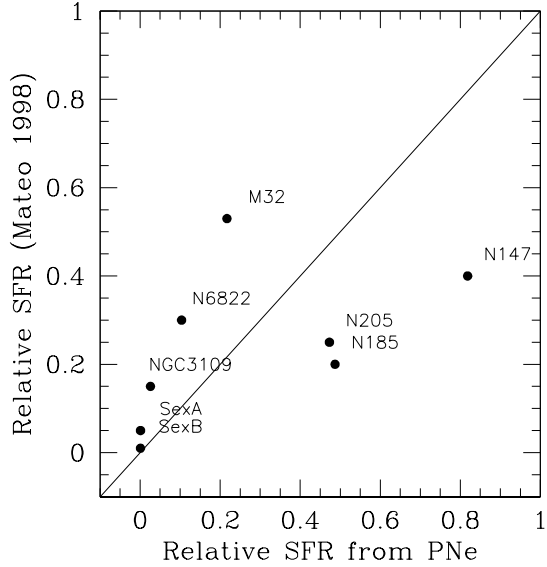}
\caption{The relative SFR in the intermediate-age for LG galaxies: computed from the number of PNe  
extrapolated to 4~mag below the bright end of the PNLF (x axes) and the SFR from  Mateo (1998) (y axes).}
\end{figure}

The PN population size, when corrected for the
different completeness limits of the various surveys, scales 
well with the V luminosity of a galaxy (\cite{m03b},\cite{cm05}), 
but it is expected to be sensitive to the relative amount 
of star formation during the epoch of the formation of the PN progenitors,  the
last 1 to 8 Gyr (the so called intermediate-age). 
In fact, their number reflects the total mass of the underlying stellar population 
from which they derive.
The total mass of the progenitor population can be derived 
using the theoretical relation between the total luminosity of the 
population of PNe progenitors and
the PNe number, obtained under the hypothesis of coeval, chemically
homogeneous stars (\cite{rb}).  
The total luminosity of the
population of PNe progenitors, $L_{T_{\rm pPNe}}$, is related to the
number of PNe, $n_{\rm PNe}$, to their lifetime, $t_{\rm PNe}$, and to
the so-called evolutionary flux, $\dot{\xi}_{\rm PNe}$.  The
number of stars with initial mass $0.8M_{\odot}<M<8 M_{\odot}$ per unit luminosity
leaving the Main Sequence each year is then given by the expression
\begin{equation} 
L_{T_{\rm pPNe}}=\frac{n_{\rm PNe}} {t_{\rm PNe} \dot{\xi}_{\rm PNe}}. 
\end{equation}  

Thus, counting planetary nebulae, extrapolating their number to the
same completeness limit, i.e. 4~mag below the bright end of the PNLF,
and estimating the mean mass of the PNe progenitors, we have a measure
of the total mass of the intermediate-age stellar population
(\cite{m05b}).  The intermediate-age stellar mass is thus proportional
to the number of PNe: for each PN observed a total mass of
$\sim$2$\times$10$^6$~M$_{\odot}$ composed by stars with initial mass
$0.8M_{\odot}<M<8 M_{\odot}$ is formed.  Considering the whole IMF
(i.e. \cite{scalo98}) we have that 10$^7$~M$_{\odot}$ were produced for each PNe
observed.
In Figure~1, we show the SFR known from the literature
(cf. \cite{mateo98}), vs. that  computed with PNe, from the ratio of the
mass of intermediate-age population and the total stellar mass.  The
solid line is the bisector of the first quadrant.  A good agreement
between the two SFRs can be noticed.  The most remarkable exception is
NGC147, where the SFR derived from PNe is higher than expected.  PNe
are therefore confirmed to be useful evolutionary age-tracers of the
intermediate-age population. The presence of PNe is a good and
valuable evidence for an intermediate-age population
(cf. \cite{apariciogallart94}).  Moreover the relationship of their
number to the host galaxy mass gives quantitative information on the
relative star formation rate.

\subsection{Spectroscopy} 
 
The elemental abundances measured in PNe are extremely important to
investigate the metal content of the host galaxy a few Gyrs ago, at
the time when their stellar progenitors formed.  In fact, the PN
progenitor do not process elements like S and Ar, and to a first
approximation also Ne and O; therefore these abundances in PNe probe
the metallicity of the interstellar medium when their progenitor stars
were born.  On the other hand, the progenitors of PNe produce and
return to the interstellar medium elements like He, C, and N.
Therefore, measuring abundances of these elements allows to test
stellar evolution models.  
The PNe in the LG have known
distances, and thus the luminosity and temperature of their central
stars can be estimated via modelling with CLOUDY, as done in
Magrini et al. (2005A).  The masses of the PN central stars, and hence the mass
and age of their progenitors, can then be calculated from the central
stars location on the Hertzprung-Russell diagram.  
Completing the
study of the samples of PNe with HII regions allows us to plot the
age-metallicity relationship and thus to study the chemical enrichment
of the host galaxy (see Figure~2).

As an example of this kind of study, we report the results obtained with the 
VLT spectroscopy of the PN and HII region populations in Sextans A and Sextans B (\cite{m05a}). 
The main results are:
\begin{itemize}  
\item[{\it i)}] The O/H abundance of HII regions is  homogeneous throughout the
galaxies,  contrary to the large dishomogenities  expected due to the episodic 
star formation and to the solid body rotation of dIr galaxies (\cite{s89}).  

\item[{\it ii)}] The age-metallicity relation is almost constant for both 
galaxies, indicating a very low (or absent)
chemical enrichment in these galaxy in the last 10 Gyr (see
Figure~2).

\item[{\it iii)}] The present time O/H abundances from HII regions  of Sextans~A and Sextans~B are comparable 
within the errors, 12+log(O/H)=7.6$\pm$0.2 for Sextans~A and 7.8$\pm$0.2 for 
Sextans B (\cite{m05a}), in spite of quite different star formation histories. 
This result confirms that galaxies with similar luminosity and morphological type have 
also similar metallicity, and maintain it over a large time interval. 
This means that the well-known metallicity-luminosity relationship represents 
the ability of a galaxy to keep the product of its own evolution more than to produce metals. 

\item[{\it iv)}]In low metallicity stars the nucleosynthesis and dredge-up processes are
predicted to be more efficient than in solar metallicity stars
(i.e. \cite{herwig04},\cite{marigo01}), but observations of Galactic PNe to test the
theoretical models are still missing.  
The oxygen overabundance of the only PN known in Sextans A
compared with that of HII regions in the same galaxy, suggests oxygen
enrichment due to the third-dredge-up process  for massive (M$>$3M$_{\odot}$), 
low-metallicity PNe progenitors. This is in good agreement with the predictions 
of the above quoted theories (\cite{mar04}).

These results show how PNe and HII regions are of great importance to
set constraints to chemical evolution models of dwarf galaxies (see
above i, ii, iii) (cf. Hensler et. \cite{hensler99}, Recchi et
al. \cite{recchi01}) and also to study the nucleosynthesis of
low-metallicity stars (iv).
\end{itemize}
 
\begin{figure}
\label{fig-sex}
\centering
 \includegraphics[height=13cm,width=8cm,angle=-90]{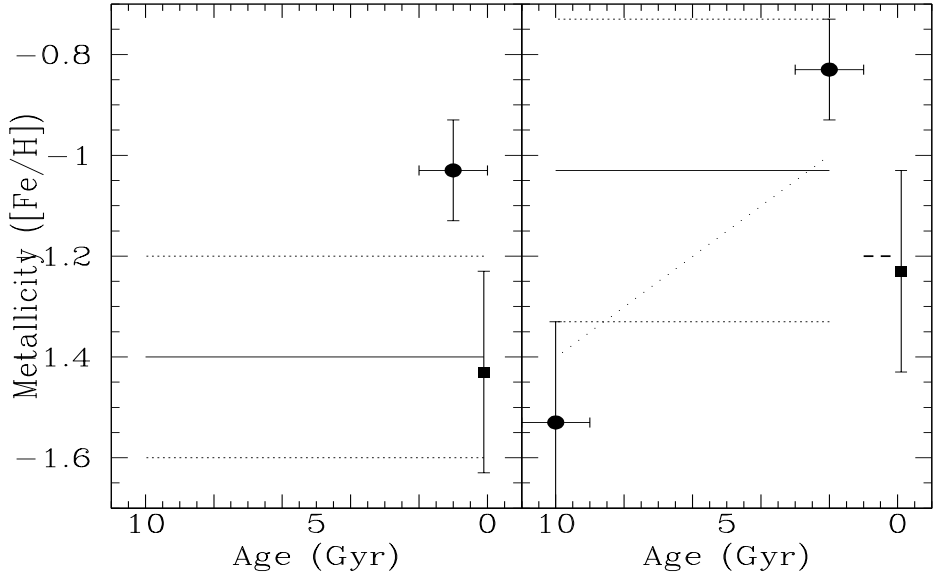}
\caption{The age-metallicity relationship for the dIr galaxies Sextans A 
(left panel) and Sextans B (right panel).  For Sextans A: the solid
line is the mean [Fe/H] of stars derived from colour-magnitude
diagrams (CMD) by Dolphin et al. (\cite{d03}), while the dotted lines
represent the spread of stellar metallicities.  The square is the mean
of metallicity of the HII regions and the error bars are the rms
scatter, the filled circle is the  PN abundance.  
For Sextans B: the solid is the mean abundance of the Sextans B PNe
with its rms scatter (dotted lines), the filled circles are the
abundances of the oldest and the youngest PNe.  The square is the mean
metallicity of the HII regions and the error bars are the rms
scatter. 
The dashed line is the stellar metallicity derived by
CMD by Tosi et al. (\cite{t91}).  The transformation from [O/H] to
[Fe/H] needed to compare our nebular oxygen abundances with the
stellar abundances was computed according to the empirical
transformation by Mateo (1998): [Fe/H]=[O/H]-0.37, where
[O/H]=$\log$~(O/H)-$\log$~(O/H)$_{\odot}$.  We used the solar
abundances from Asplund (\cite{asplund03}) to derive [O/H] from the
measured $12 + \log~$(O/H). }

\end{figure}

{\bf Acknowledgments}:\\ 
I would thank the SOC of this Conference for having invited me and for the IAU 
grant.
It is a pleasure to thanks the LGC team for the work done together during these years.


\begin{thebibliography}{} 

\addcontentsline{toc}{section}{References} 
\bibitem[Aparicio \& Gallart 1994]{apariciogallart94}
Aparicio A., Gallart C.,  1994,
in  The Local Group: Comparative and Global Properties, ESO Conference and Workshop Proceedings,  Eds Andrew Layden, 
R. Chris Smith, and Jesper Storm, p.115
 
\bibitem[Arnaboldi et al. 2003]{arna2003}  Arnaboldi, M., Freeman, K. C., Okamura, S., et al. 2003, AJ, 125, 514 

\bibitem[2003]{asplund03}  Asplund, M. 2003, 
CNO in the Universe, ed. C. Charbonnel, D. Schaerer,  G. Meynet (San Francisco: ASP), ASP Conf. Ser., 304, 275 

\bibitem[Baade (1955)]{b55} Baade, W., 1955, AJ, 60, 151  
 
\bibitem[Cannon 1933]{cannon1933} 
Cannon, A. J., 1933, HCOB, 891, 2

\bibitem[Ciardullo et al. 1989]{c89} Ciardullo, R., Jacoby, G.H., Ford, H.C., Neill, J.D. 1989, 
ApJ, 339, 53
 
\bibitem[Ciardullo et al. 2002]{c02}   Ciardullo R., Feldmeier J.J., Jacoby G.H., Kuzio de Naray R., 
Laychak M., Durrell P.R., 2002, ApJ, 577, 31
 
\bibitem[2004]{ci04} Ciardullo, R., Durrell, P. R., Laychak, M. B., et al. 2004, ApJ, 614, 167  

\bibitem[Corradi \& Magrini 2005]{cm05} Corradi, R.L.M., Magrini, L., 2005, in the proceedings of the ESO-Workshop "Planetary Nebulae beyond the Milky Way", Eds. L. Stanghellini, J.R. Walsh, N.G. Douglas,  p.36 

\bibitem[Corradi et al. 2005]{c05} Corradi, R.L.M., Magrini, L., Greimel, R., et al. 2005, A\&A, 
431, 555
 
\bibitem[Corradi et al. 2006]{c06} Corradi, R.L.M., Magrini, L., Leisy, P., et al. 2006, in preparation
 

\bibitem[Danziger et al. 1978]{d78} Danziger, I.J., Webster, B.L., Dopita, M.A., Hawarden, 
T.G, 1978, ApJ, 220, 458 
 
\bibitem[2003]{d03} Dolphin, A. E., Saha, A., Skillman, E. D., et al., 2003, 
AJ, 126, 187
 
\bibitem[Douglas et al. 2002]{douglas02} Douglas, N. G.; Arnaboldi, M.; Freeman, K. C., et al.,  
PASP, 114,  801, 1234 
 
\bibitem[Feldmeier et al. 2003]{f03} Feldmeier, J. J., Ciardullo, R., Jacoby, G. H., \& Durrell, P. R.,
2003, ApJS, 145, 65 

\bibitem[Ford et al. 1973]{f73} Ford, H.C., Jenner, D.C., Epps, H.W., 1973, AJ, 183, 73 
 
\bibitem[Ford et al. 1977]{f77} Ford, H.C., Jacoby, G.H., Jenner, D.C., 1977, ApJ, 213, 
18
 
\bibitem[Ford \& Jacoby 1978]{fj78} Ford, H.C., Jacoby, G.H. 1978, ApJS, 38, 351 

\bibitem[Gon\c{c}alves et al. 2006]{g06} Gon\c{c}alves, R.D., Magrini, L., Leisy, P., Corradi, R., 2006, this book

\bibitem[1999]{hensler99} Hensler, G., Rieschick, A., Koepen, J.,1999, ASPC, 187, 214

\bibitem[Hernandez \& Pe\~{n}a 2006]{her06} Hernandez-Martinez, L., Pe\~{n}a, M., 2006, this book

\bibitem[Herwig 2004]{herwig04}
Herwig, F., 2004, 
ApJS, 155, 651  

\bibitem[Jacoby 1980]{j80} Jacoby, G.H., 1980, ApJS, 42, 1
 
\bibitem[1989]{j89} Jacoby, G.H., 1989, ApJ,  339, 39 

\bibitem[Jacoby 2005]{j05} Jacoby, G.H., 2005, in the proceedings of the ESO-Workshop "Planetary Nebulae beyond the Milky Way", Eds. L. Stanghellini, J.R. Walsh, N.G. Douglas,  p.17


\bibitem[Jacoby \& Ciardullo 1999]{jc99} Jacoby, G. H., Ciardullo, R., 1999, ApJ, 515, 169
 
\bibitem[Jacoby \& Lesser 1981]{jl81} Jacoby, G.H., Lesser, M.P., 1981, AJ, 86, 185  

\bibitem[Leisy \& Dennefeld 2006A]{l06a} Leisy, P., Dennefeld, M., 2006, A\&A in press

\bibitem[Leisy et al. 2006B]{l06b} Leisy, P., Magrini, L., Corradi, R., 2006, in preparation

\bibitem[Leisy et al. 2005]{l05} Leisy, P., Corradi, R.L.M., Magrini, L., Greimel, R., 
Mampaso, A., Dennefeld, M., 2004, A\&A, 436, 437 
 
\bibitem[1987]{l87} Lequeux, J., Meyssonnier, N., Azzopardi, M., 1987, 
A\&AS, 67, 169

\bibitem[Lindsay (1955)]{l55}Lindsay, E.M., 1955, MNRAS, 115, 248


\bibitem[2000]{m00} Magrini L., Corradi, R.L.M., Mampaso, A., Perinotto, 
M. 2000, A\&A, 355, 713
 
\bibitem[Magrini et al. 2002]{m02} Magrini, L., Corradi, R.L.M., Walton, N.A., Zijlstra, 
A. A., Pollacco, D.L., Walsh, J.R., Perinotto, M., Lennon, D.J.,
Greimel, R., 2002, A\&A, 386, 869


\bibitem[Magrini et al. 2003A]{m03a} Magrini, L., Perinotto, M., Corradi, R. L. M., Mampaso, A., 2003, 
A\&A, 400, 511
 
\bibitem[Magrini et al. 2003B]{m03b} Magrini, L., Corradi, R.L.M., Greimel, R., Leisy, P., 
Lennon, D.J., Mampaso, A., Perinotto, M., Pollacco, D.L., Walsh, J.R.,
Walton, N.A., Zijlstra, A.A., 2003, A\&A, 407, 51

\bibitem[Magrini et al. 2005A]{m05a} Magrini, L., Leisy, P., Corradi, R.L.M., Perinotto, 
M., Mampaso, A., Vilchez, J.M.,  2005A, A\&A, 443, 115 
 
\bibitem[Magrini et al. 2005B]{m05b} Magrini, L., Corradi, R.L.M., Greimel, R., et al., 2005B, MNRAS, 361, 517

\bibitem[Magrini et al. 2006]{m06} Magrini, L., Corradi, R.L.M., Leisy, P., 2006, in prep. 

\bibitem[Marigo 2001]{marigo01} 
Marigo, P., 2001, 
A\&A, 370, 194 
 
\bibitem[Marigo et al. 2004]{mar04} Marigo, P., Girardi, L., Weiss, A., Groenewegen, 
M.A.T., Chiosi, C., 2004, A\&A, submitted
 
\bibitem[Mateo  1998]{mateo98}  Mateo, M., 1998	A\&A Rev 36, 435 
 
\bibitem[Massey et al. 2002]{massey02} Massey, P., Hodge, P.W., Holmes, S., et al., 2002,  
AAS, 201, Vol. 34, p.1272
 
\bibitem[Merrett et al. 2006]{merrett06} Merrett, H. R., Merrifield, M. R., Douglas, N.G., et al., 2006,  astro-ph/0603125 

\bibitem[Parker 2006]{p06} Parker, Q.A., this book


\bibitem[Pe\~{n}a et al. 2006]{pena06} Pe\~{n}a, M., Richer, M., Stasinska, G., 2006, this book
 
\bibitem[2001]{recchi01} Recchi, S., Matteucci, F.,  D'Ercole, A,, 2001, MNRAS, 322, 800

\bibitem[Reid \& Parker 2005]{rp05} Reid, W. \& Parker, Q.A., in the proceedings of the ESO-Workshop "Planetary Nebulae beyond the Milky Way", Eds. L. Stanghellini, J.R. Walsh, N.G. Douglas,  p.30

\bibitem[Renzini \& Buzzoni 1986]{rb} Renzini, A., Buzzoni, A., 1986 in {\em Spectral 
Evolution of Galaxies}, eds. Chiosi C. and Renzini A., Reidel,
Ap. Space Sci. Lib., Vol. 122, p.195
  
\bibitem[Richer \& McCall 2002]{r02} 
Richer, M. G., McCall, M. L., 2002, RMxAC, 12, 173

\bibitem[Scalo 1998]{scalo98} 
Scalo J.M., 1998, 
in The Stellar Initial Mass Function, ed, G. Gilmore, \& D. Howell (San Francisco:ASP), 
ASP Conf. Ser., 142, 201
     
\bibitem[Skillman et al. 1989]{s89} Skillman, E. D., Kennicutt, R. C., Hodge, P. W., 1989, 
ApJ, 347, 875

\bibitem[1991]{t91} Tosi, M., Greggio, L., Marconi, G., \& Forcardi, P. 1991, 
AJ, 102, 951

\bibitem[van den Bergh 2000]{vdb00} van den Bergh, S., 2000, in {\em The Galaxies of the 
Local Group}, (Cambridge: Cambridge University Press)
 
\bibitem[van Zee et al. 2006]{vz06} van Zee, L., Skillman, E. D., Haynes, M. P.,
2006, ApJ, 637, 269

\bibitem[1968]{w68} Westerlund, B.E., IAU Symp. n. 34, Osterbrock D.E., 
O'Dell C.R. eds., Dordrecht, p.23 

\bibitem[Zijlstra et al. 2006]{z06} Zijlstra, A., Gesicki, K., Walsh, J. R., Pequignot, D., 
van Hoof, P. A. M., Minniti,  D., MNRAS accepted, astro-ph/0603422

\end{thebibliography}
\end{document}